\documentclass[prc,twocolumn,amsmath,amssymb,floatfix,nofootinbib]{revtex4}
\usepackage{mathrsfs,bm}
\usepackage{longtable,lscape}
\usepackage{txfonts}
\usepackage{indentfirst}
\usepackage{graphicx,color,dcolumn,booktabs}
\usepackage{multirow}
\usepackage{indentfirst}
\usepackage{multirow}
\usepackage{epsfig}
\usepackage{graphicx,color,dcolumn}
\usepackage{epstopdf}
\usepackage{amsmath}
\usepackage{multirow}
\usepackage{diagbox}
\usepackage{changes}
\usepackage{threeparttable}

\usepackage{color}
\usepackage{amssymb}
\definecolor{cover}{rgb}{0.77,0.87,0.88}
\definecolor{blueone}{rgb}{0.1,0.1,.7}
\definecolor{citec}{rgb}{0.14,0.47,0.09}
\definecolor{two}{rgb}{0.0,0.5,0.}
\definecolor{three}{rgb}{.5,.1,0.15}
\usepackage[bookmarks=true,bookmarksopen=false,plainpages=false,breaklinks=true,
   bookmarksnumbered=true,hypertexnames=false,
   filecolor=blue,urlcolor=three,menucolor=three,
   linkcolor=three,citecolor=blueone, colorlinks,
   anchorcolor=blue,runcolor=pink,frenchlinks=red
   pdfstartview=FitH,pdftitle=title,%
   pdfauthor=author]{hyperref}

\usepackage{relsize}
\usepackage{xspace}
\def\babar{\mbox{\slshape B\kern-0.1em{\smaller A}\kern-0.1em
    B\kern-0.1em{\smaller A\kern-0.2em R}}}

\allowdisplaybreaks[4]
\usepackage{color}
\begin{document}
\title{Possible molecular states from  interactions of charmed baryons}

\author{Dan Song, Lin-Qing Song, Shu-Yi Kong, Jun He\footnote{Corresponding author: junhe@njnu.edu.cn}}
\affiliation{School of Physics and Technology, Nanjing Normal University, Nanjing 210097, China}

\date{\today}
\begin{abstract}

In this work,  we perform a systematic study of  possible molecular states
composed of two charmed baryons including  hidden-charm systems $\Lambda_c
\bar{\Lambda}_c$, $\Sigma_c^{(*)}\bar{\Sigma}_c^{(*)}$, and $\Lambda_c
\bar{\Sigma}_c^{(*)}$, and corresponding  double-charm systems
$\Lambda_c\Lambda_c$, $\Sigma_c^{(*)}\Sigma_c^{(*)}$, and $\Lambda_c
\Sigma_c^{(*)}$. With the help of the heavy quark chiral effective Lagrangians,
the interactions are described with  $\pi$, $\rho$,  $\eta$, $\omega$, $\phi$,
and $\sigma$ exchanges. The potential kernels are constructed, and inserted into
the quasipotential Bethe-Salpeter equation. The bound states from the
interactions considered is studied by searching for the poles of the scattering
amplitude. The results suggest that strong attractions exist in both hidden-charm and
double-charm systems considered in the current work, and bound states can be
produced in most of the systems. More experimental studies about
these molecular states are suggested though the nucleon-nucleon collision at LHC and
nucleon-antinucleon collision  at $\rm \bar{P}ANDA$.

\end{abstract}

\maketitle

\section{Introduction}

With the development of the experimental technology, a large amount of data
accumulated in experiment provide opportunity to the study of the hadron spectrum.
In the recent years, more and more hadrons have been observed in
experiment~\cite{PDG}.  Many of these new observed hadrons cannot be put into
the conventional  quark model, which is the basic frame to understand the hadron
spectrum~\cite{Godfrey:1985xj,Capstick:1986ter}. A growing number of efforts have been
paid to explain their origin and internal structure. An obvious observation is
that many newly observed particles are close to the threshold of  two hadrons,
so a popular picture to understand these exotic hadrons is the molecular state,
which is a loosely bound state of hadrons. The $XYZ$ particles, such as
$X(3872)$, $Z_c(3900)$ and $Z_b(10610)$ and $Z_c(10650)$, were widely assigned
as molecular states in the
literature~\cite{Chen:2016qju,Guo:2017jvc,Tornqvist:1993ng,He:2014nya,Sun:2011uh}.
Particularly, the observed hidden-charm pentaquarks provide a wonderful spectrum
of molecular states composed of an anticharmed meson and a charmed
baryon~\cite{Wu:2010jy,Yang:2011wz,Xiao:2013yca,Chen:2015loa,He:2015cea,Liu:2019tjn,He:2019ify}.
Such picture is enhanced by the recent observed strange hidden-charm
pentaquarks~\cite{Peng:2020hql,Wang:2020eep,Chen:2016ryt,Zhu:2021lhd}. However,
though the well-known deuteron and the dibaryon with nucleon, $\Delta$, and
$\Lambda$ baryon were predicted and studied in both theory and experiment very
far before the XYZ particle and pentaquarks, few predicted molecular states of two
baryons are observed in experiment~\cite{Clement:2016vnl,PDG}.  Some theoretical
studies have been performed to discuss the possibility of existence of molecular
states composed of two baryons beyond nucleon, $\Delta$, and $\Lambda$
baryon~\cite{vanBeveren:2008rt,Zhu:2019ibc,Lee:2011rka,Dong:2021juy,Song:2022yfr,Liu:2011xc}.

Most of the molecular state candidates observed in the past two decades are in
the hidden-charm sector. Hence, it is natural to expect the molecular state
composed of a charmed baryon and an anticharmed baryon. In  recent years, the
structures near the $\Lambda_c\bar{\Lambda}_c$ threshold has attracted much
attentions. A charmoniumlike $Y(4630)$ with quantum numbers $J^{PC}=1^{--}$ was
observed at Belle~\cite{Belle:2008xmh}. After the experimental discovery of
$Y(4630)$,  many theoretical works have performed to understand its origin, such
as conventional charmonium state~\cite{Badalian:2008dv,Segovia:2008ta} and
compact multiquark
state~\cite{Maiani:2014aja,Cotugno:2009ys,Brodsky:2014xia,Wang:2021qmn}. Due to
the closeness of the mass of $Y(4630)$ and the $\Lambda_c\bar{\Lambda}_c$ threshold,
the relation between $Y(4630)$ and the threshold effect was studies in
Ref.~\cite{vanBeveren:2008rt}.  In Ref.~\cite{Simonov:2011jc}, the mechanism of
$Y(4630)$ enhancement in $\Lambda_c\bar \Lambda_c$ electroproduction was also
studied.  The $\Lambda_c\bar{\Lambda}_c$  molecular state also attracts much
attention~\cite{Chen:2011cta,Wang:2021qmn,Lee:2011rka,Simonov:2011jc}.
Theoretical calculations suggest strong attraction between a $\Lambda_c$ baryon
and an $\bar{\Lambda}_c$ baryon by $\sigma$ and $\omega$ exchanges, which favors
the existence of a $\Lambda_c \bar{\Lambda}_c$ molecular
state~\cite{Lee:2011rka,Simonov:2011jc}. In our previous work, the $\Lambda_c
\bar{\Lambda}_c$ molecular state can be produced from the interaction, but it is
difficult to be used to interpret the $Y(4630)$~\cite{Song:2022yfr}. The
studies of more molecular states with a charmed baryon and an anticharmed
baryon are also helpful to understand this exotic structure. In the current work, the
interactions $\Lambda_c\bar{\Lambda}_c$, $\Sigma_c^{(*)}\bar{\Sigma}_c^{(*)}$,
and $\Lambda_c \bar{\Sigma}_c^{(*)}$ will be studied in a quasipotential
Bethe-Salpeter equation (qBSE) approach.

In our model, the double-charm molecular states can be obtained by replacing the
anticharmed hadron by a charmed
hadron~\cite{Ding:2020dio,Ding:2021igr,Kong:2021ohg}. The recent experimental
observation exhibits the ability to observe double-charm hadrons in experiment.
The LHC Collaboration reported a state $\Xi_{cc}^{++}$ ~\cite{LHCb:2017iph},
which indicts the possibility of experimental observation of double-heavy
molecular state. Very recently, the LHCb Collaboration observed an open charm
tetraquark state $T_{cc}^{+}$ below the $D^0D^{*+}$ mass
threshold~\cite{LHCb:2021vvq}, which has already been predicted by a lot of
theoretical works in the diquark and antidiquark picture
~\cite{Ader:1981db,Zouzou:1986qh,Lipkin:1986dw,Heller:1986bt,Carlson:1987hh,Silvestre-Brac:1993zem,Semay:1994ht,Gelman:2002wf},
also in the molecular picture
~\cite{Manohar:1992nd,Pepin:1996id,Molina:2010tx,Li:2012ss,Wang:2017uld,Maiani:2019lpu,Liu:2019stu}.
The doubly charm dibaryon attracts some attentions from the
hadron physics community~\cite{Li:2012bt,Garcilazo:2020acl,Carames:2015sya,Chen:2021cfl,Ling:2021asz,Dong:2021bvy}.
Hence, in the current work, the double-charm systems $\Lambda_c\Lambda_c$,
$\Sigma_c^{(*)}\Sigma_c^{(*)}$, and $\Lambda_c \Sigma_c^{(*)}$ will be also
calculated.

This article is organized as follows. After the Introduction, Section~\ref{Sec:
Formalism} shows the details of dynamics of the charmed baryons interactions,
reduction of potential kernel and a brief introduction of the qBSE. In
Section~\ref{Sec:Results and discussions}, the numerical results are
given.  Finally, summary  and  discussion are given in
Section~\ref{Sec:Summary}.

\section{Theoretical frame}\label{Sec: Formalism}

To  study the interactions of charmed baryons, we need to construct the
potential kernel, which is performed by introducing the exchanges of
peseudoscalar $\mathbb{P}$, vector $\mathbb{V}$ and scalar $\sigma$ mesons. The
Lagrangians depicting the couplings of light mesons and  baryons are required and
will be presented  below.

\subsection{Relevant Lagrangians}

The Lagrangians for the couplings between charmed baryon and light mesons are constructed under the heavy quark limit and  chiral symmetry as~\cite{Liu:2011xc,Isola:2003fh,Falk:1992cx},
\begin{align}
{\cal L}_{S}&=-
\frac{3}{2}g_1(v_\kappa)\epsilon^{\mu\nu\lambda\kappa}{\rm tr}[\bar{S}_\mu
{\cal A}_\nu S_\lambda]+i\beta_S{\rm tr}[\bar{S}_\mu v_\alpha (\mathcal{V}^\alpha-
\rho^\alpha)
S^\mu]\nonumber\\
& + \lambda_S{\rm tr}[\bar{S}_\mu F^{\mu\nu}S_\nu]
+\ell_S{\rm tr}[\bar{S}_\mu \sigma S^\mu],\nonumber\\
{\cal L}_{B_{\bar{3}}}&= i\beta_B{\rm tr}[\bar{B}_{\bar{3}}v_\mu(\mathcal{V}^\mu-\rho^\mu)
B_{\bar{3}}]
+\ell_B{\rm tr}[\bar{B}_{\bar{3}}{\sigma} B_{\bar{3}}], \nonumber\\
{\cal L}_{int}&=ig_4 {\rm tr}[\bar{S}^\mu {\cal A}_\mu B_{\bar{3}}]+i\lambda_I \epsilon^{\mu\nu\lambda\kappa}v_\mu{\rm tr}[\bar{S}_\nu F_{\lambda\kappa} B_{\bar{3}}]+H.c.,
\end{align}
where $S^{\mu}_{ab}$ is composed of the Dirac spinor operators,
\begin{align}
    S^{ab}_{\mu}&=-\sqrt{\frac{1}{3}}(\gamma_{\mu}+v_{\mu})
    \gamma^{5}B^{ab}+B^{*ab}_{\mu}\equiv{ B}^{ab}_{0\mu}+B^{ab}_{1\mu},\nonumber\\
    \bar{S}^{ab}_{\mu}&=\sqrt{\frac{1}{3}}\bar{B}^{ab}
    \gamma^{5}(\gamma_{\mu}+v_{\mu})+\bar{B}^{*ab}_{\mu}\equiv{\bar{B}}^{ab}_{0\mu}+\bar{B}^{ab}_{1\mu},
\end{align}
and the bottomed baryon matrices are defined as
\begin{align}
B_{\bar{3}}&=\left(\begin{array}{ccc}
0&\Lambda^+_c&\Xi_c^+\\
-\Lambda_c^+&0&\Xi_c^0\\
-\Xi^+_c&-\Xi_c^0&0
\end{array}\right),\quad
B=\left(\begin{array}{ccc}
\Sigma_c^{++}&\frac{1}{\sqrt{2}}\Sigma^+_c&\frac{1}{\sqrt{2}}\Xi'^+_c\\
\frac{1}{\sqrt{2}}\Sigma^+_c&\Sigma_c^0&\frac{1}{\sqrt{2}}\Xi'^0_c\\
\frac{1}{\sqrt{2}}\Xi'^+_c&\frac{1}{\sqrt{2}}\Xi'^0_c&\Omega^0_c
\end{array}\right). \nonumber\\
B^*&=\left(\begin{array}{ccc}
\Sigma_c^{*++}&\frac{1}{\sqrt{2}}\Sigma^{*+}_c&\frac{1}{\sqrt{2}}\Xi^{*+}_c\\
\frac{1}{\sqrt{2}}\Sigma^{*+}_c&\Sigma_c^{*0}&\frac{1}{\sqrt{2}}\Xi^{*0}_c\\
\frac{1}{\sqrt{2}}\Xi^{*+}_c&\frac{1}{\sqrt{2}}\Xi^{*0}_c&\Omega^{*0}_c
\end{array}\right).\label{MBB}
\end{align}

The explicit forms of the Lagrangians can be written as,
\begin{align}
{\cal L}_{BB\mathbb{P}}&=-i\frac{3g_1}{4f_\pi\sqrt{m_{\bar{B}}m_{B}}}~\epsilon^{\mu\nu\lambda\kappa}\partial^\nu \mathbb{P}~
\sum_{i=0,1}\bar{B}_{i\mu} \overleftrightarrow{\partial}_\kappa B_{j\lambda},\nonumber\\
{\cal L}_{BB\mathbb{V}}&=-\frac{\beta_S g_V}{2\sqrt{2m_{\bar{B}}m_{B}}}\mathbb{V}^\nu
 \sum_{i=0,1}\bar{B}_i^\mu \overleftrightarrow{\partial}_\nu B_{j\mu}\nonumber\\
&-\frac{\lambda_S
g_V}{\sqrt{2}}(\partial_\mu \mathbb{V}_\nu-\partial_\nu \mathbb{V}_\mu) \sum_{i=0,1}\bar{B}_i^\mu B_j^\nu,\nonumber\\
{\cal L}_{BB\sigma}&=\ell_S\sigma\sum_{i=0,1}\bar{B}_i^\mu B_{j\mu},\nonumber\\
    {\cal L}_{B_{\bar{3}}B_{\bar{3}}\mathbb{V}}&=-\frac{g_V\beta_B}{2\sqrt{2m_{\bar{B}_{\bar{3}}}m_{B_{\bar{3}}}} }\mathbb{V}^\mu\bar{B}_{\bar{3}}\overleftrightarrow{\partial}_\mu B_{\bar{3}},\nonumber\\
{\cal L}_{B_{\bar{3}}B_{\bar{3}}\sigma}&=i\ell_B \sigma \bar{B}_{\bar{3}}B_{\bar{3}},\nonumber\\
{\cal L}_{BB_{\bar{3}}\mathbb{P}}
    &=-i\frac{g_4}{f_\pi} \sum_i\bar{B}_i^\mu \partial_\mu \mathbb{P} B_{\bar{3}}+{\rm H.c.},\nonumber\\
{\cal L}_{BB_{\bar{3}}\mathbb{V}}    &=\frac{g_\mathbb{V}\lambda_I}{\sqrt{2m_{\bar{B}}m_{B_{\bar{3}}}}} \epsilon^{\mu\nu\lambda\kappa} \partial_\lambda \mathbb{V}_\kappa\sum_i\bar{B}_{i\nu} \overleftrightarrow{\partial}_\mu
   B_{\bar{3}}+{\rm H.c.}.
   \label{LB}
\end{align}
The $\mathbb{V}$ and $\mathbb{P}$ are the vector and pseudoscalar matrices as
\begin{align}
{\mathbb P}=\left(\begin{array}{ccc}
        \frac{\sqrt{3}\pi^0+\eta}{\sqrt{6}}&\pi^+&K^+\\
        \pi^-&\frac{-\sqrt{3}\pi^0+\eta}{\sqrt{6}}&K^0\\
        K^-&\bar{K}^0&\frac{-2\eta}{\sqrt{6}}
\end{array}\right),
\mathbb{V}=\left(\begin{array}{ccc}
\frac{\rho^{0}+\omega}{\sqrt{2}}&\rho^{+}&K^{*+}\\
\rho^{-}&\frac{-\rho^{0}+\omega}{\sqrt{2}}&K^{*0}\\
K^{*-}&\bar{K}^{*0}&\phi
\end{array}\right).
\end{align}
The masses of  particles involved in the calculation are chosen as suggested central values in the Review of  Particle Physics  (PDG)~\cite{PDG}. The mass of broad $\sigma$ meson is chosen as 500 MeV.  The  coupling constants involved are listed in Table~\ref{coupling}.
\renewcommand\tabcolsep{0.13cm}
\renewcommand{\arraystretch}{1.5}
\begin{table}[h!]
\caption{The coupling constants adopted in the
calculation, which are cited from the literature~\cite{Chen:2019asm,Liu:2011xc,Isola:2003fh,Falk:1992cx,Zhu:2022fyb,Zhu:2021lhd}. The $\lambda$ and $\lambda_{S,I}$ are in the units of GeV$^{-1}$. Others are in the units of $1$.
\label{coupling}}
\begin{tabular}{cccccccccccccccccc}\bottomrule[1pt]
$\beta$&$g$&$g_V$&$\lambda$ &$g_{s}$\\
0.9&0.59&5.9&0.56 &0.76\\\hline
$\beta_S$&$\ell_S$&$g_1$&$\lambda_S$ &$\beta_B$&$\ell_B$ &$g_4$&$\lambda_I$\\
-1.74&6.2&-0.94&-3.31&$-\beta_S/2$&$-\ell_S/2$&$3g_1/{(2\sqrt{2})}$&$-\lambda_S/\sqrt{8}$ \\
\bottomrule[1pt]
\end{tabular}
\end{table}

First, we should construct flavor wave functions with definite isospin under
$SU(3)$ symmetry. In this paper, we take the following charge conjugation
conventions for two-baryon system as ~\cite{Dong:2021juy},
\begin{equation}
|B_1B_2\rangle_c=\frac{1}{\sqrt{2}}|B_1\bar{B}_2-(-1)^{J-J_1-J_2}cc_1c_2|B_2\bar{B}_1\rangle,
\end{equation}
where $J$ and $J_i$ are the spins of system $|B_1\bar{B}_2\rangle$ and $|B_i\rangle$,  respectively, and $c_i$ is defined by $\mathcal{C}|B_i\rangle = c_i\bar{B}_i\rangle$. For the isovector state, the $C$ parity cannot be defined, so we will use the $G$ parity  instead as $G=(-1)^{I}C$ with $C=c$.

Following the method in Ref.~\cite{He:2019rva}, we input vertices $\Gamma$ and  propagators $P$  into the code directly.  The potential can be written as
\begin{equation}%
{\cal V}_{\mathbb{P},\sigma}=I_{\mathbb{P},\sigma}\Gamma_1\Gamma_2 P_{\mathbb{P},\sigma}f(q^2),\ \
{\cal V}_{\mathbb{V}}=I_{\mathbb{V}}\Gamma_{1\mu}\Gamma_{2\nu}  P^{\mu\nu}_{\mathbb{V}}f(q^2).\label{V}
\end{equation}

In this work, both hidden-charm and double-charm systems will be considered in the calculation. The well-known $G$-parity rule will be adopted to write the interaction of a charmed and an anticharmed baryon from the interaction of two charmed baryons. By inserting the $G^{-1}G$ operator into the
potential, the $G$-parity rule can be obtained easily
as~\cite{Phillips:1967zza,Klempt:2002ap,Lee:2011rka,Zhu:2019ibc},
\begin{eqnarray}
V&=&\sum_{i}{\zeta_{i}V_{ihh}}.
\end{eqnarray}
The $G$ parity of the exchanged meson is left as a $\zeta_{i}$ factor for $i$ meson.

The propagators are defined as usual as
\begin{equation}%
P_{\mathbb{P},\sigma}= \frac{i}{q^2-m_{\mathbb{P},\sigma}^2},\ \
P^{\mu\nu}_\mathbb{V}=i\frac{-g^{\mu\nu}+q^\mu q^\nu/m^2_{\mathbb{V}}}{q^2-m_\mathbb{V}^2},
\end{equation}
where the form factor $f(q^2)$ is adopted to compensate the off-shell effect of exchanged meson as $f(q^2)=e^{-(m_e^2-q^2)^2/\Lambda_e^2}$
with $m_e$ and $q$ being the $m_{\mathbb{P},\mathbb{V},\sigma}$ and the momentum of the exchanged  meson.
The $I_i$ is the flavor factor for certain meson exchange $i$ of certain interaction, and the explicit values are listed in Table~\ref{flavor factor}.
\renewcommand\tabcolsep{0.19cm}
\renewcommand{\arraystretch}{2}
\begin{table}[h!]
\begin{center}
\caption{The flavor factors $I_i^d$  and $(-1)^{(J+1)} I_i^c$ of exchange $i$ for direct diagram and cross diagram, respectively. The values in bracket are for the heavy-heavy baryons if the values are different from these of heavy-antiheavy baryons.}
\label{flavor factor}
\begin{tabular}{c|c|ccccc}\bottomrule[1pt]
  $I_i^d$ &$I$&$\pi$&$\eta$  &$\rho$ &$\omega$&$\sigma$  \\\hline
$\Lambda_c \bar{\Lambda}_c[\Lambda_c \Lambda_c]$&$0$&$ $&$ $&$ $ &$-2[2]$ & $4[4]$\\
$\Sigma_c^{(*)}  \bar{\Sigma}_c^{(*)}[\Sigma_c^{(*)}\Sigma_c^{(*)}]$&$0$&$1[-1]$&$\frac{1}{6}[\frac{1}{6}]$&$-1[-1]$&$-\frac{1}{2}[\frac{1}{2}]$& $1[1]$\\
$ $&$1$&$\frac{1}{2}[-\frac{1}{2}]$&$\frac{1}{6}[\frac{1}{6}]$&$-\frac{1}{2}[-\frac{1}{2}]$&$-\frac{1}{2}[\frac{1}{2}]$& $1[1]$\\
$ $&$2$&$-\frac{1}{2}[\frac{1}{2}]$&$\frac{1}{6}[\frac{1}{6}]$&$\frac{1}{2}[\frac{1}{2}]$&$-\frac{1}{2}[\frac{1}{2}]$& $1[1]$\\
$\Lambda_c \bar{\Sigma}_c^{(*)}[\Lambda_c \Sigma_c^{(*)}]$&$1$&$ $&$ $&$ $&$-1[1]$& $2[2]$\\
\bottomrule[1pt]
  $(-1)^{(J+1)} I_i^c$  &$I$&$\pi$&$\eta$  &$\rho$ &$\omega$ \\\hline
$\Lambda_c \bar{\Sigma}_c[\Lambda_c \Sigma_c]$&$1$&$c[1]$&$ $&$-c[1]$ &$ $  \\
$\Lambda_c \bar{\Sigma}_c^*[\Lambda_c \Sigma_c^*]$&$1$&$-c[1]$&$ $&$c[1]$ &$ $ \\
$\Sigma_c  \bar{\Sigma}_c^*[\Sigma_c  \Sigma_c^*]$&$0$&$c[-1]$&$\frac{c}{6}[\frac{1}{6}]$&$-c[-1]$&$-\frac{c}{2}[\frac{1}{2}]$ \\
$ $&$1$&$\frac{c}{2}[-\frac{1}{2}] $&$\frac{c}{6}[\frac{1}{6}] $&$-\frac{c}{2}[-\frac{1}{2}] $ &$-\frac{c}{2}[\frac{1}{2}] $  \\
$ $&$2$&$-\frac{c}{2}[\frac{1}{2}] $&$ \frac{c}{6}[\frac{1}{6}]$&$\frac{c}{2}[\frac{1}{2}] $ &$-\frac{c}{2}[\frac{1}{2}] $  \\
\toprule[1pt]
\end{tabular}
\end{center}
\end{table}

With the potential kernel obtained, the qBSE is adopted to solve the scattering
amplitude~\cite{He:2014nya,He:2015mja,He:2012zd,He:2015yva,He:2017aps,Zhu:2021lhd,Kong:2021ohg,Ding:2020dio}.
The 4-dimensional Bethe-Saltpeter equation in the Minkowski space can be reduced
to a 1-dimensional  equation with fixed spin-parity $J^P$
as~\cite{He:2015mja}, after partial-wave decomposition and spectator
quasipotential approximation.
\begin{align}
i{\cal M}^{J^P}_{\lambda'\lambda}({\rm p}',{\rm p})
&=i{\cal V}^{J^P}_{\lambda',\lambda}({\rm p}',{\rm
p})+\sum_{\lambda''}\int\frac{{\rm
p}''^2d{\rm p}''}{(2\pi)^3}\nonumber\\
&\cdot
i{\cal V}^{J^P}_{\lambda'\lambda''}({\rm p}',{\rm p}'')
G_0({\rm p}'')i{\cal M}^{J^P}_{\lambda''\lambda}({\rm p}'',{\rm
p}),\quad\quad \label{Eq: BS_PWA}
\end{align}
where the sum extends only over non-negative helicity $\lambda''$.
The $G_0({\rm p}'')$ is reduced from the 4-dimensional  propagator $G(p'')$ under quasipotential approximation with one of two baryons on-shell as \begin{align}
	G(p'')&=\frac{\delta^+(p''^{~2}_h-m_h^{2})}{p''^{~2}_l-m_l^{2}}
	\nonumber\\
	\to G_0({\rm p}'')&=\frac{1}{2E_h({\rm p''})[(W-E_h({\rm
p}''))^2-E_l^{2}({\rm p}'')]}.
\end{align}
where $p''_l$ and $m_l$ are the momentum and mass of light hadron, respectively. 
As required by the spectator approximation, the heavier particle is on shell, which satisfies  $p''^0_h=E_{h}({\rm p}'')=\sqrt{
m_{h}^{~2}+\rm p''^2}$. The $p''^0_l$ for the lighter particle is then $W-E_{h}({\rm p}'')$. Here and hereafter, a definition ${\rm p}=|{\bm p}|$ will be adopted. 

The partial wave potential is defined with the potential of  interaction obtained in the above in Eq.~(\ref{V}) as
\begin{align}
{\cal V}_{\lambda'\lambda}^{J^P}({\rm p}',{\rm p})
&=2\pi\int d\cos\theta
~[d^{J}_{\lambda\lambda'}(\theta)
{\cal V}_{\lambda'\lambda}({\bm p}',{\bm p})\nonumber\\
&+\eta d^{J}_{-\lambda\lambda'}(\theta)
{\cal V}_{\lambda'-\lambda}({\bm p}',{\bm p})],
\end{align}
where $\eta=PP_1P_2(-1)^{J-J_1-J_2}$ with $P$ and $J$ being parity and spin for system. The initial and final relative momenta are chosen as ${\bm p}=(0,0,{\rm p})$  and ${\bm p}'=({\rm p}'\sin\theta,0,{\rm p}'\cos\theta)$. The $d^J_{\lambda\lambda'}(\theta)$ is the Wigner d-matrix.
We also adopt an  exponential regularization  by introducing a form factor into the propagator as
$G_0({\rm p}'')\to G_0({\rm p}'')\left[e^{-(p''^2_l-m_l^2)^2/\Lambda_r^4}\right]^2$ with $\Lambda_r$ being a cutoff~\cite{He:2015mja}.

\section{Numerical Results}\label{Sec:Results and discussions}

With the  preparation above, numerical calculation can be performed to study the
molecular states from the interactions $\Lambda_c
\bar{\Lambda}_c/\Lambda_c\Lambda_c$,
$\Sigma_c^{(*)}\bar{\Sigma}_c^{(*)}/\Sigma_c^{(*)}\Sigma_c^{(*)}$, and
$\Lambda_c \bar{\Sigma}_c^{(*)}/\Lambda_c \Sigma_c^{(*)}$. After transformation
of the one dimensional integral qBSE into a matrix equation, the scattering
amplitude can be obtained, and the molecular states can be searched for as the
poles of the amplitude. The parameters of the Lagrangians in the current work
are chosen as the same as those in our previous study of the hidden-charm
pentaquarks~\cite{He:2019ify,He:2019rva}. The only free parameters are cutoffs
$\Lambda_e$ and $\Lambda_r$, which are rewritten as a form of
$\Lambda_r=\Lambda_e=m+\alpha~0.22$ GeV with $m$ being the mass of the exchanged meson, which is also introduced into the regularization form factor to suppress
large momentum, i. e., the short-range contribution of
the exchange as warned in Ref~\cite{Liu:2019zvb}. Hence, in the current work, only one
parameter $\alpha$ is involved.

\subsection{Interactions $\Lambda_c \bar{\Lambda}_c$ and $\Lambda_c\Lambda_c$  }

\begin{figure}[h!]
  \centering
 \includegraphics[scale=0.62,bb=90 350 500 530,clip]{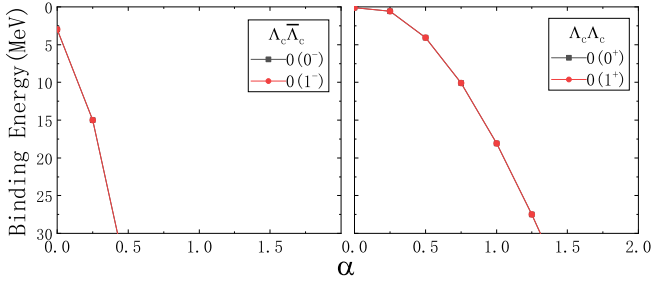}\\
  \caption{The  binding energies of the bound states from $\Lambda_c \bar{\Lambda}_c$ (left) and $\Lambda_c\Lambda_c$ (right) interactions with the variation of parameter $\alpha$. }\label{0}
\end{figure}

In the current work, only S-wave states will be considered. For the two
interactions considered, the results with spins $S$=1 and 0  are shown in
Fig.~\ref{0}. The results suggest bound states are produced from all four
channels. The states with spins 1 and 0 have almost the same binding energy,
which is consistent with the results in Ref.~\cite{Lee:2011rka}. The two bound
states from the $\Lambda_c \bar{\Lambda}_c$  interaction appear even with an
$\alpha$ value below 0, which are smaller than two states for the double-charm
$\Lambda_c \Lambda_c$ interaction, which indicts that the $\Lambda_c
\bar{\Lambda}_c$ interaction is more attractive than the $\Lambda_c \Lambda_c$
interaction due to different contributions from the meson exchanges. Since the
$\Lambda$ baryon is isoscalar, the interactions $\Lambda_c \bar{\Lambda}_c$ and
$\Lambda_c\Lambda_c$ arises from the $\sigma$ and $\omega$ exchanges. In the
$\Lambda_c \bar{\Lambda}_c$ interaction, both $\sigma$ and $\omega$ exchanges
provide attraction.  However, in the $\Lambda_c\Lambda_c$ interaction, the
$\omega$ exchange is repulsive, which reduces the attraction.

\subsection{Interactions $\Sigma_c^{(*)}\bar{\Sigma}_c^{(*)}$ and $\Sigma_c^{(*)}\Sigma_c^{(*)}$  }

\begin{figure}[h!]
  \centering
  \includegraphics[scale=0.62,bb=90 50 500 530,clip]{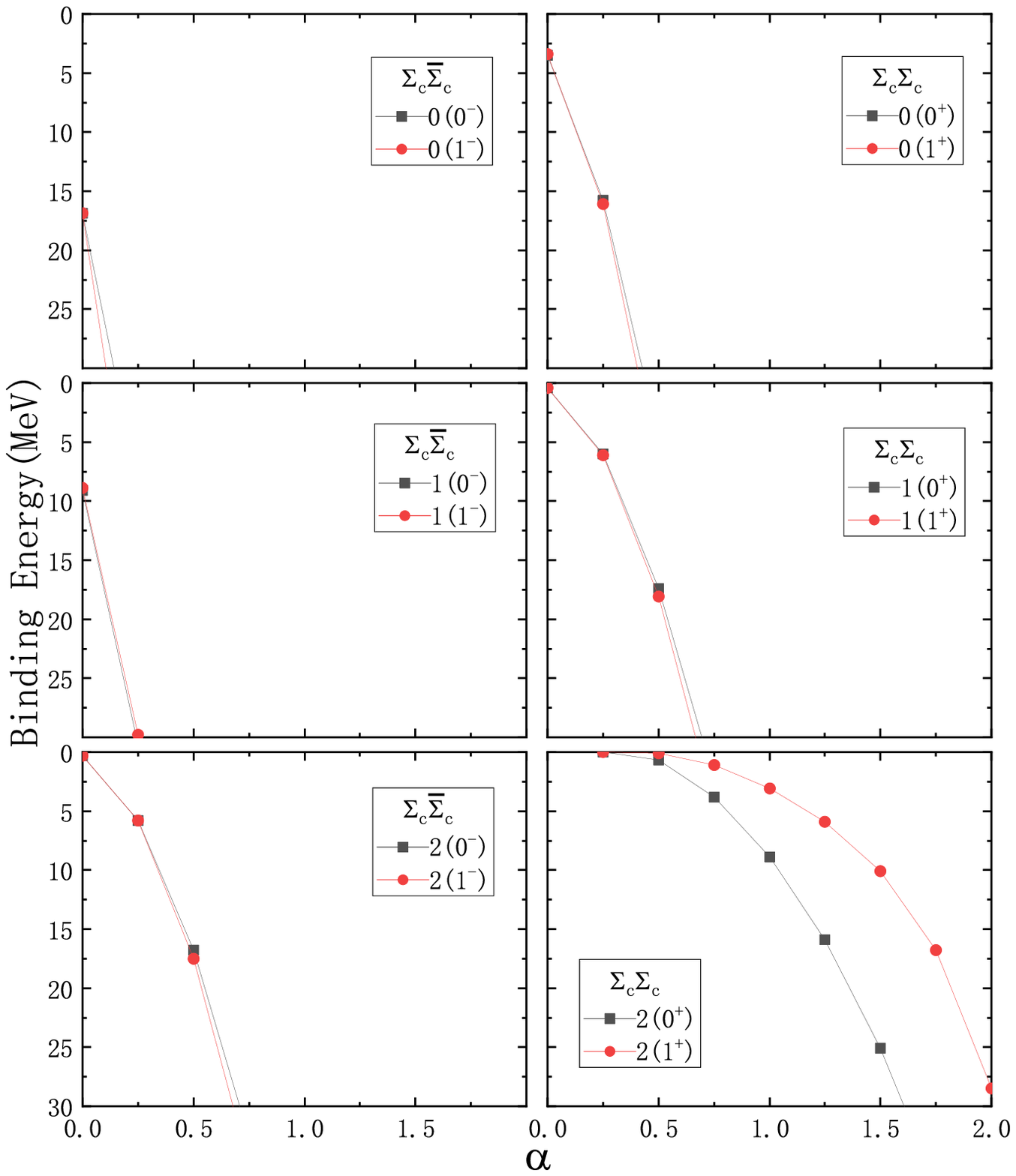}\\
  \caption{The  binding energies of the bound states from $\Sigma_c \bar{\Sigma}_c$ (left) and $\Sigma_c\Sigma_c$ (right) interactions with the variation of parameter $\alpha$.}\label{scsc}
\end{figure}

Different from the isoscalar $\Lambda_c$ baryon, the $\Sigma_c^{(*)}$ baryon is an isovector particle. Hence, more channels will be involved in certain interaction. In Fig.~\ref{scsc}, the bound states from the $\Sigma_c \bar{\Sigma}_c$ interaction and their double-charm partners are presented. Here, the siospin $I$ can be 0, $1$, or $2$, and the spin $S$=0 or $1$, which leads to six channels for each interaction.

As shown in Fig.~\ref{scsc}, bound states are produced in all channels, but with
different behaviors with the variation of parameter $\alpha$. For the isoscalar
hidden-charm $\Sigma_c\bar{\Sigma}_c$ system, the bound states are produced at
an $\alpha$ value below 0, and the binding energies increase rapidly with the
increase of $\alpha$ value. As shown in Table~\ref{flavor factor}, the strong
attraction is from the $\rho$ exchange with a large flavor factor $-1$.  The
corresponding double-charm partners appear at larger $\alpha$ value, which means
that it is less attractive than the hidden-charm case due to the different signs
for $\pi$ and $\omega$ exchanges. The binding energies for states with different spins are almost the same. For the states with $I=1$, the
binding energies at an $\alpha$ value of 0 are smaller than those with $I=0$.
As shown in Table~\ref{flavor factor}, the flavor factors for $\rho$ and $\pi$
exchanges are half of those for $I=0$, which leads to less
attraction.   For the states with $I=2$, the attraction becomes weaker due to
reversing the signs of the $\rho$ and $\pi$ exchanges.  The hidden-charm
states are produced at a small $\alpha$ value, and binding energies increase to
a value larger than 30 GeV very quickly at an $\alpha$ value of about 0.7. However,
the binding energies of their double-charm partners appear at $\alpha$ value of
about 0.2, and increase relatively slowly. 

\begin{figure}[h!]
  \centering
    \includegraphics[scale=0.62,bb=90 50 500 530,clip]{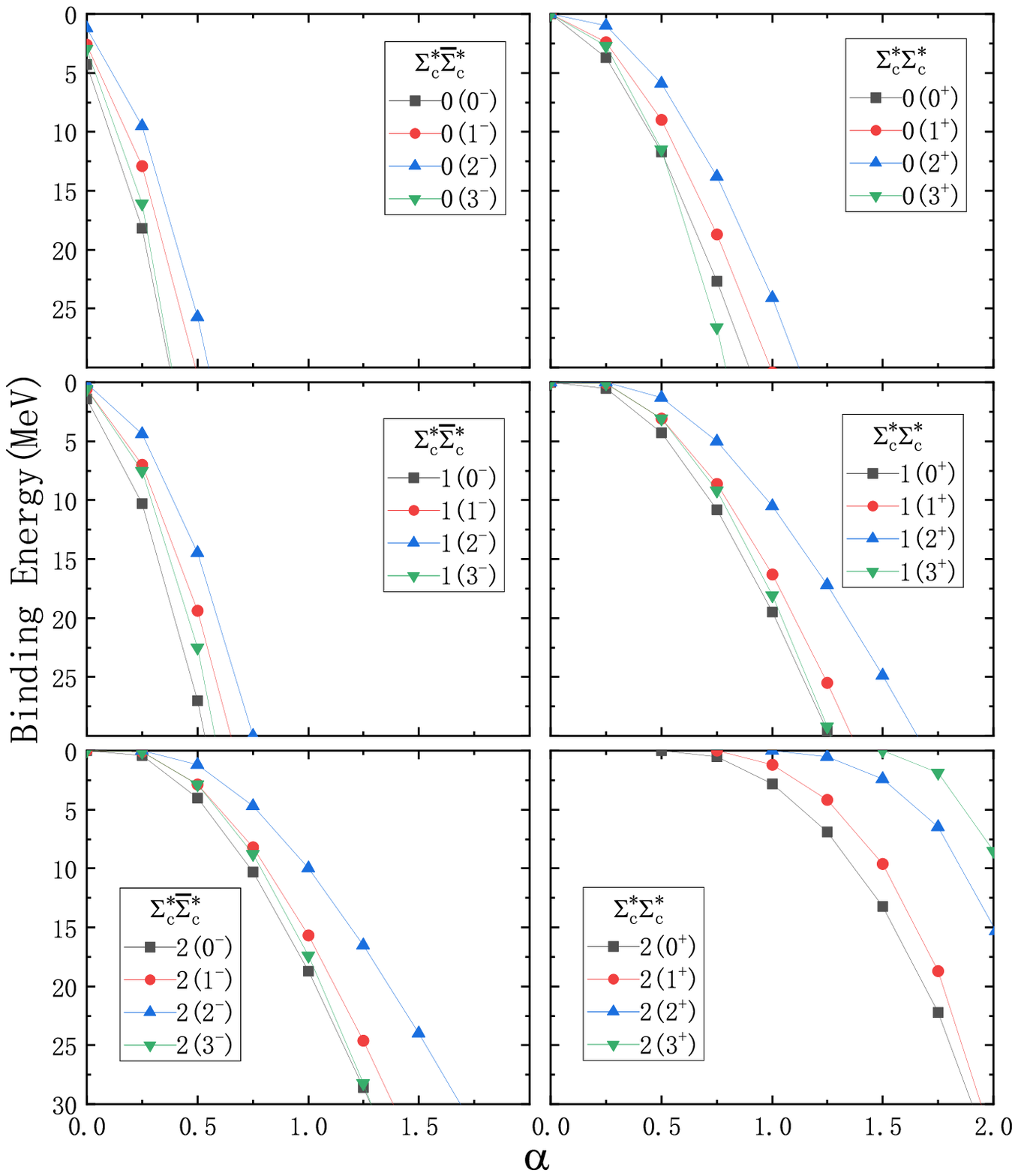}\\
  \caption{The  binding energies of the bound states from $\Sigma_c^* \bar{\Sigma}_c^*$ (left) and $\Sigma_c^* \Sigma_c^* $ (right) interactions with the variation of parameter $\alpha$. }\label{scasca}
\end{figure}

The binding energies of the states produced from the $\Sigma_c^*
\bar{\Sigma}_c^*$ interaction are shown in Fig.~\ref{scasca}. 
Except that there
are four spins $S$=0, 1, 2, and 3, due to the flavor factors are the same as
those for the $\Sigma_c\bar{\Sigma}_c$ system, the results are similar to the
results in Fig.~\ref{scsc}.  For the hidden-charm system with $I=0$, there are
three states with spins $J=1$, 2, and 3 producing at an $\alpha$ values of about
0.  As in the case of $\Sigma_c\bar{\Sigma}_c$, the attractions for the
corresponding double-charm systems are weaker than the hidden-charm systems. The
hidden-charm bound states with $I=1$ appear at an $\alpha$ value of about 0, and
the binding energies increase to 30 MeV at $\alpha$ value about 0.7. The
hidden-charm states with $I=2$ appear at an $\alpha$ value little larger than 0
while their double-charm partners appear at $\alpha$ value of 0.5 or larger.
Generally speaking, the attractions of $\Sigma_c^* \bar{\Sigma}_c^*/\Sigma_c^*
{\Sigma}_c^*$ interaction are a little weaker than the case of $\Sigma_c
\bar{\Sigma}_c/\Sigma_c\Sigma_c$ interaction.

\begin{figure}[h!]
  \centering
  \includegraphics[scale=0.62,bb=90 60 500 530,clip]{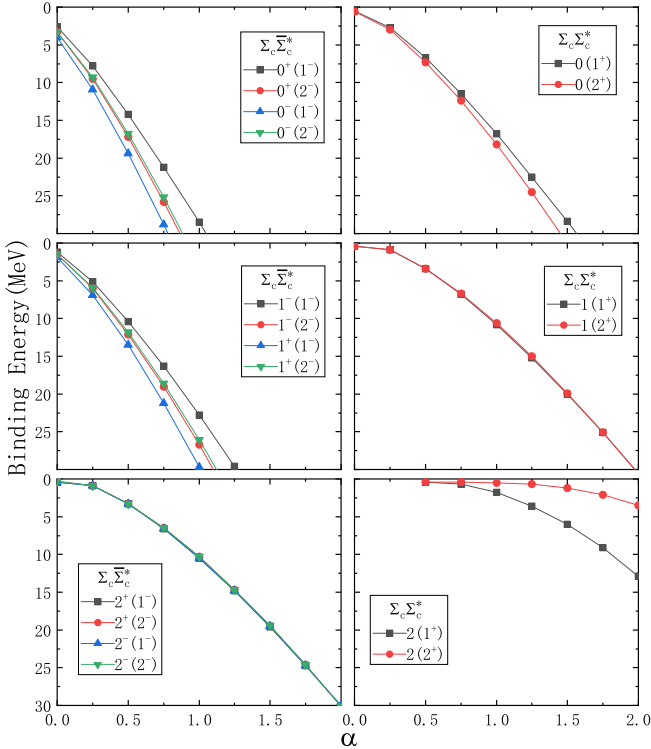}\\
  \caption{The  binding energies of the bound states from $\Sigma_c \bar{\Sigma}_c^*$ (left) and $\Sigma_c\Sigma_c^*$ (right) interactions with the variation of parameter $\alpha$. }\label{scsca}
\end{figure}

The results for the $\Sigma_c \bar{\Sigma}_c^*$ and $\Sigma_c\Sigma_c^*$
interactions are presented in Fig~\ref{scsca}.  For the hidden-charm states,
there are two $G$ parities, $G=\pm1$, which do not involve in the double-charm
sector. For the hidden-charm systems with $I$=0, the bound states appear at
$\alpha$ value a little below 0, and increase with the increase of the parameter
$\alpha$ to 30 MeV at $\alpha$ value of about 1. For their double-charm partners,
the bound states appear at an $\alpha$ value  about 0, and the binding
energies increase more slowly than the hidden-charm states. In the case with
$I$=1, the states appear at an $\alpha$ value of about 0, and increase to 30 MeV
at an $\alpha$ value about 1.2.  The hidden-charm states with $I=2$ appear at
$\alpha$ value of about 0, which is smaller than these for the double-charm
states, about 0.5.

\subsection{Interactions $\Lambda_c \bar{\Sigma}_c^{(*)}$ and $\Lambda_c \Sigma_c^{(*)}$  }

\begin{figure}[h!]
  \centering
  \includegraphics[scale=0.62,bb=90 210 500 530,clip]{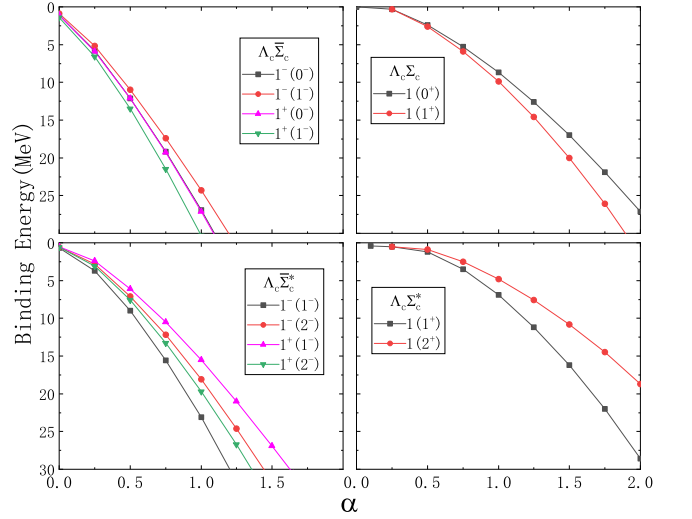}\\
  \caption{The  binding energies of the bound states from $\Lambda_c\bar{\Sigma}_c$ (upper left) and $\Lambda_c\Sigma_c $ (upper right) $\Lambda_c\bar{\Sigma}_c^*$ (bottom left) and $\Lambda_c\Sigma_c^*$ (bottom right)  interactions with the variation of parameter $\alpha$. }\label{lcsc}
\end{figure}

Now, we turn to the systems with a $\Lambda_c/\bar{\Lambda}_c$ and a
$\Sigma^{(*)}$ baryon. The results of the $\Lambda_c\bar{\Sigma}_c^{(*)}$ states
and their double-charm partners are shown in Fig.~\ref{lcsc}. Here, the S-wave
states with spin $J=0$ and 1 are considered. Since the
$\Lambda_c/\bar{\Lambda}_c$ baryon is isoscalar, the isospin only can be 1, and
the $G$ parity will involve in the hidden charm sector. Due to the same flavor
factors, the results of systems with $\Sigma_c$ and $\Sigma_c^*$ are similar.
The hidden-charm states are first produced at an $\alpha$ value a little below
0, and the binding energies increase to 30 GeV at an $\alpha$ value about 1. The
double-charm states appear at an $\alpha$ value a litter larger and the
binding energies increase slowly, reach 30 MeV at an $\alpha$ value of  about 2.

\section{Summary}\label{Sec:Summary}

In the current work, the study of the molecular states from interactions of
charmed baryons is performed. The hidden-charm systems $\Lambda_c
\bar{\Lambda}_c$,  $\Sigma_c^{(*)}\bar{\Sigma}_c^{(*)}$, and $\Lambda_c
\Sigma_c^{(*)}$, as well as their double-charm partners, are considered in the
calculation. With the help of the Lagrangians in heavy quark limit and with
chiral symmetry. The potential kernels are constructed in a
one-boson-exchange model, and inserted into the qBSE to search the bound states.

The calculation suggests that the attractions widely exist in the systems of two
charmed baryons. For the $\Lambda_c\bar{\Lambda}_c$ interaction, the bound
states are produced with spin parities $J^P$=$0^-$ and $1^-$, and their
double-charm partner can be produced with a binding energies smaller than 30 MeV
in a larger range of the parameter $\alpha$. Due to the same favor factors for
the $\Sigma_c\bar{\Sigma}_c$, $\Sigma_c^{*}\bar{\Sigma}_c^{*}$, and
$\Sigma_c\bar{\Sigma}_c^{*}$ interactions, the binding energies for these three
interactions behave in a similar manner. The most strong attraction can be found
in the case with $I=0$ for both hidden-charm and doubly-charm cases due to the
large $\rho$ exchange as suggested by its flavor factor, which is consistent with the results in Ref.~\cite{Dong:2021juy,Dong:2021bvy}. For the interactions
$\Lambda_c \bar{\Sigma}_c^{(*)}$ and $\Lambda_c \Sigma_c^{(*)}$, all bound
states produced are relatively stable, has a binding energy below 30 ~MeV in a
large range of $\alpha$ value. Generally speak, the interactions of two charmed
baryons are attractive, and many bound states are produced.  However, only a few
candidates, such as $Y(4630)$, were reported in experiment.  More experiment
studies about these states are suggested though the processes including the
nucleon- nucleon collision at LHC and nucleon-antinucleon collision  at $\rm
\bar{P}ANDA$.

\noindent {\bf Acknowledgement} We would like to thank Prof. Feng-Kun Guo for
helpful discussions. This project is supported by the National Natural Science
Foundation of China (Grants No. 11675228).

\end{document}